# Youthful perspectives on sustainability: Examining pro-environmental behaviors in tourism through latent class cluster analysis

Riccardo Gianluigi Serio[1], Maria Michela Dickson[2*], Giuseppe Espa[1] and Rocco Micciolo[3]

[1]Department of Economics and Management, University of Trento, Italy
[2]Department of Statistical Sciences, University of Padova, Italy
[3]Department of Psychology and Cognitive sciences, University of Trento, Italy

**Abstract**
Tourism has emerged as a significant driver of the global economy. As its economic impact grows, concerns regarding environmental sustainability have intensified. This paper explores the dual dimensions of sustainable tourism: the relationship between tourism supply and sustainability, and tourist demand characteristics. It highlights the critical role of young tourists, who exhibit a heightened awareness of environmental issues and advocate for sustainable practices. By conducting a survey among young Italian university students, the study identifies distinct segments based on family background, political orientation, and travel habits. Utilizing latent class cluster analysis, the findings aim to enhance understanding of pro-environmental behaviors among youth, offering insights for policymakers to foster sustainable tourism practices.

___________________________

*Corresponding author: mariamichela.dickson@unipd.it

**Keywords:** Young tourists; Sustainable tourism; Latent class cluster analysis; Pro-environmental behavior.

# 1. Introduction

Tourism is a driving sector of the modern economy and has experienced a profound expansion in recent decades. Although hard-to-predict events, such as the COVID-19 pandemic that slowed tourism growth in 2020- 2021, the World Travel and Tourism Council (WTTC, 2022) indicates that tourism contributes 7.6% of global GDP. Similarly, the labor market also benefits from the effects of the tourism sector, which created 22 million new jobs worldwide in 2022 (WTTC, 2022). As the tourism sector became a more significant part of the economy, concerns about its impact on the environment began to arise.

The issue of environmental sustainability in general has become of paramount importance nowadays, especially in human activities that can induce a progressive deterioration of natural resources, such as, among others, tourism (Weaver, 2014). The WTO provided a widely accepted definition of sustainable tourism in 1993, stating it as "*Tourism that meets the current needs of tourists and host regions while preserving and enhancing opportunities for the future*". Various actions have been taken by governments and institutions to address the negative impact of tourism on the environment, including the implementation of the European Charter for Sustainable Development (Europarc Federation, 2010) and the ETIS toolkit (ETIS, 2017). This problem has been approached in literature from two sides, one focused on exploring the relationship between tourism supply and sustainability (e.g., Massidda and Etzo, 2012; Xu and Dong, 2020), and another investigating the characteristics of the demand for tourism and tourists (e.g., Ashraf *et al*., 2020; Han, 2021). The latter state that tourists' behavior can significantly reduce the environmental impacts of tourism (Xu and Dong, 2020; Wang *et al*., 2020; Wu *et al*., 2021), achieving sustainability in the sector (Bridges and Wilhelm, 2008; Garvey and Bolton, 2017). Following this road, it seems clear that of great importance in the environment preservation of tourist destinations is the tourists' point of view and their attitude in implementing green practices. Analyzing tourists means identifying groups of them with common characteristics, so that they correctly identify demand segments to define long-term destination strategies (Dolnicar, 2004) and to understand destination's environmental sustainability dynamics.

An interesting focus can be made on a specific subgroup of tourists, namely young people. Since the attention to the environment is a relatively recent issue, younger age groups are those more involved in its preservation. The new generations exhibit a heightened awareness of environmental degradation and advocate for the imperative shift towards sustainable practices in human endeavors, aligning with long-term sustainability goals. Moreover, the engagement of young individuals in environmental protection assumes paramount importance, serving dual purposes: firstly, they play a pivotal role in comprehending the future trajectory of sustainable tourism trends; secondly, they represent the cohort upon whom proactive measures must be directed to instigate a shift towards more sustainable consumption patterns among future generations, thereby fostering environmental sustainability within the tourism sector (Pendergast, 2009; Vukic *et al*., 2015).

The present paper focuses on young tourists and endeavors to explore their inclination toward pro-environmental behavior based on individual characteristics, family background, social context, and travel habits. The main goal of the present research is to find segmentation patterns over the youngster that can ensure higher commitment towards the reduction of tourism environmental footprint. To investigate the phenomenon, a survey was implemented on young Italian students at university level. Through latent class cluster analysis, this study delineates two distinct segments of young tourists, characterized by varying family backgrounds, political orientations, and travel habits.

The paper is structured as follows. A review of the relevant literature on sustainable tourism and the pro-environmental behavior especially of young tourists is presented in Section 2, while Section 3 introduces the implemented statistical methodology. The data used for the classification are discussed in Section 4 and results are presented in Section 5, providing valuable insights to contribute to the existing literature and offering suggestions for policymakers. Concluding remarks in Section 6 summarizes key findings and proposes avenues for future research.

## 2. Literature background on sustainable tourism

Given the economic and social importance of the tourism sector globally and given the magnitude of the environmental impacts resulting from this massive industry, sustainable tourism is an increasingly studied topic by researchers (e.g., Butler, 1999; Fayos-Solà *et al*., 2014; Lu and Nepal, 2009). However, despite a considerable number of indicators have been offered in literature to drive improvements in environmental sustainability in tourism, the actual implementation of these indicators and the resulting outcome has been poor (Madhavan and Rastogi, 2013). It means that the successful transition of tourism towards sustainability entails a multifaceted process encompassing diverse dimensions, including the natural environment, economic viability, and societal welfare, alongside various individual-related aspects such as external, philosophical, and internal challenges (Streimikiene *et al*., 2021). It is thus clear that the individuals play a crucial role in the protection of the environment, by means of sustainable consumption behaviors (Wang *et al.*, 2020). In this framework, Han (2021) reports the terms *environmentally responsible behaviors*, *environmentally sustainable behaviors*, *environmentally protective/preserving behaviors*, *ecological behaviors*, and *green behaviors* as referable to the same expected behavior toward this direction. Tourists impacted by such new consciousness also prompted tour operators capable of providing green experiences (Wang *et al.*, 2018). Several studies have delved into the behavioral and motivational aspects of the *sustainable tourist*, identifying at least four distinct segments: the ecotourist (e.g., Beall and Boley, 2022), nature-based tourist (e.g., Fredman *et al.*, 2012), responsible tourist (e.g., Goodwin, 2016), and green tourist (e.g., Dolnicar and Matus, 2008). In studies focusing on tourism demand, particularly within the realm of ecotourism, Blamey (2001) emphasizes nature, education, and sustainability as primary criteria for delineating the ecotourist segment. Similarly, Khanra *et al.* (2021) underscore the significance of sustainable behavior and reflections on the carbon footprint in travel as pivotal areas warranting further research. Important contributions on the subject have also been made by Weaver and Lawton (2007), who distinguishes hard from soft ecotourists according to the degree of motivation that drives their behavior and to the willingness to pay premium prices to support sustainable travel practices, showing that true ecotourists are rare, as well as concentrated in wealthier countries. However, the defined categories lack precise delineation, exhibiting both behavioral and attitudinal overlaps, along with individual characteristics such as age, occupation, and education (Buffa, 2015).

A growing attention is recently directed toward young individuals as both current and prospective consumers of tourism services and products (Leask *et al.*, 2015). Within the tourism sector, researchers (e.g., Pendergast, 2009; Williams *et al.*, 2010) have dedicated efforts to studying Generation Y (namely, born between 1981 and 1996,) and Generation Z (namely, born after 1997). Despite ongoing debate concerning differences among the two, these younger cohorts share defining characteristics that markedly differentiate them from previous generations. As noted by Moscardo *et al.* (2011), the emerging generations display distinctive skills, behaviors, and values, possessing enhanced communication abilities, largely proficiency in the English language, and enjoying rapid access to a vast array of information through readily available technology (Yeoman *et al.*, 2010). Furthermore, they represent the first cohort to primarily socialize online (Reinikainen *et al.*, 2020), generally boast higher levels of education compared to preceding generations (Moscardo *et al.*, 2011), possessing abundant resources and discretionary time (Gardyn, 2002), and exhibiting a reduced inclination towards saving, preferring present consumption, particularly in pursuit of memorable experiences (Jennings *et al.*, 2009). All these characteristics have contributed to the formation of a generation aware of environmental sustainability issues, which in their tourist behaviors show different levels of commitment toward sustainability (Buffa, 2015). Many researchers explore the phenomenon of young ecotourists through the lens of values (Schwartz, 1994) and motivations (Fayos-Solà *et al.*, 2014; Gillison *et al*., 2019), focusing on behavioral aspects and investigating the personality traits of young individuals or on perceptions of sustainability (Sánchez-Fernández *et al*., 2019).

## 3. Latent class cluster analysis

To explore patterns of varying propensities for environmental sustainability, a cluster analysis method, namely Latent Class Cluster Analysis (LCA), needs to be employed. LCA operates as a probabilistic algorithm, presuming that the data originate from a multivariate distribution (or mixed-modes data, as termed by Everitt, 1988; Everitt *et al*., 1992), which models the probabilities of an observation belonging to each cluster. It is crucial to note that in LCA, membership probabilities (which dictate the number and nature of clusters) are defined post-hoc, i.e., determined by the data themselves based on whether they belong to the latent class. LCA falls within the category of mixture models, as cluster membership is not defined univocally; rather it is expressed in terms of a higher or lower probability of membership in latent classes based on manifest variables (Wedel and Kamakura, 2000). Several researchers (e.g., Rigdon *et al*., 2010) have highlighted the limitations of a priori methods compared to a posteriori method (Wedel and DeSarbo, 1994), especially when analyzing questionnaire results, where the assumptions about clusters made a priori by the researcher may not adequately capture response diversity (DeSarbo and DeSarbo, 2001). The data-driven, a posteriori nature of segments or clusters constitutes a considerable advantage in analyses involving categorical variables (i.e., surveys) because its probabilistic nature facilitates addressing the prevalent issue of heterogeneity (e.g., Kamakura and Wedel, 1995; Vermunt and Magidson, 2002), which can distort or invalidate segmentation analyses, particularly when attempting to identify patterns not directly observable (e.g., attitudes towards sustainability). Thus, through LCA, researchers can uncover unobserved and often unobservable characteristics, enabling them to offer a segmentation that accounts for the heterogeneity stemming from the unobservability of classes themselves. Finally, a further advantage of LCA over other segmentation methods is that it provides an objective and formal measure for determining the exact number of clusters. Using the Bayesian information criterion (BIC, i.e., a decreasing function of the sum of the squares of the residuals of the estimated model, expresses a measure of goodness of fit), it is easy to compare different models, assuming an increasing number of latent classes, and finally choosing the one with the lowest BIC (i.e., the model that minimizes the sum of the squares of the residuals).

Below is the formal algorithm of the latent class cluster analysis for mixed variables:

$$f(\mathbf{y}_i \mid \theta) = \sum_{k=1}^{K} \pi_k \prod_{j=1}^{J} f_k(y_{ij} \mid \theta_{jk})$$

In this formalization:
- $y_i$ represents the vector of observed variables (endogenous variables, items, or dependent variables) for observation $i$, reflecting the scores of an individual object on the observed variables. Conditional independence within latent classes is assumed.
- $\sum f(y_i|\theta)$: is the joint probability density of the sets of variables $y_{ij}$ for observation $ij$ conditional on the model parameters $\theta$. This represents the joint distribution of variables for an observation within the dataset.
- $\sum_{k=1}^{K}$ : indicates the sum over all $K$ latent classes in the model. In LCA, $K$ represents the number of $k$ unobserved latent classes or groups in the dataset.
- $\pi_k$ : is the parameter representing the a priori probability of an observation belonging to a latent class $k$. In other words, $\pi_k$ is the probability that a randomly chosen observation belongs to class $k$ before observing the data.
- $\prod_{j=1}^{J}$ : indicates the productivity on all variables $j$ in the set of variables $\mathbf{y}_i$.
- $f_k(y_{ij}|\theta_{jk})$: is the conditional probability density model for variable $j$ for the latent class $k$, which describes the distribution of the variable $y_{ij}$ for observation $i$ given membership of class $k$.
- $\theta_{jk}$ are the parameters of the model that define the distribution of the variable $y_{ij}$ for class $k$.

To summarize, the formulation describes how the joint probability density of the sets of variables for each observation can be decomposed into a weighted sum of the conditional probability densities for each latent class, where the weights are given by the a priori probabilities of belonging to the latent classes. Within each latent class, the conditional probability density is expressed as the product of the conditional probability densities for each variable within the set of variables. As anticipated, the two most important advantages of using the LCA method in survey-based studies are the ability of the model to fit non-normal distributions of the data (such as nominal or categorical variables), and the lower number of a priori assumptions required by the researcher on the number of clusters. LCA allows the data to express which number of clusters offers the best model (in terms of fit). This can be tested with methods such as likelihood-ratio tests between nested models, or measures such as the Akaike information criterion (AIC), consistent Akaike information criterion (CAIC) and BIC (Fraley and Raftery, 1998). These measures consider both the goodness of fit of the model to the data and its complexity, with more complex models being penalized; in general, a model that minimizes BIC over other models is considered preferable. To ensure the robustness of clusters obtained with LCA, particularly in the absence of access to the entire population of young tourists. However, LCA facilitates the identification of distinct cluster characteristics more effectively, particularly as the algorithm is well-suited for studies involving categorical variables.

## 4. Data description and presentation

The present study is based on a survey conducted in spring semester 2023 in the University of Trento (North-East of Italy), a middle-size university. The number of students in 2023 was more than 16 thousand, whereof more than 60% were non-resident students (and half of these came from neighbor regions). A sample of 300 students was selected by means of convenience sampling. This choice was due to privacy constraints in contacting students and following commonly utilized approaches in sustainability and segmentation literature (e.g., Kim *et al.*, 2015; Barr *et al.*, 2011). The questionnaire was composed by 14 questions, divided in three main sections. The first section delves into the level of knowledge and perception of young individuals regarding environmental sustainability within the tourism sector. The second section collects data on travel habits for holidays, expenditure patterns, and attitudes towards sustainable tourism. The final section gathers socio-demographic information, including family background, deemed pertinent for identifying discernible patterns among young people concerning travel habits, consumption behaviors, and environmental sustainability awareness. We decided to involve in the study only Italian native language students, to avoid bias due to translation or wrong understanding of the questions. Out of the administered questionnaires, 170 were completed (57% response rate).

As explicated in the introduction, the aim of the present study is to examine young tourists as present and future consumers of tourism products, aiming to discern any distinguishing characteristics among those who demonstrate a heightened awareness of environmental sustainability and a greater inclination towards pro-sustainability behaviors. This includes assessing their propensity to pay a premium for environmentally friendly travel-related activities. To achieve this goal, a two-step approach was undertaken. Firstly, following an exploratory data analysis, a cluster analysis employing Latent Class Cluster Analysis was conducted to outline segments of young tourists. Then, to assess whether there was a statistically significant association between identified clusters and propensities to adopt pro-environmental behavior, a $\chi^2$ test was employed.

**4.1 Exploratory analysis**

Table 1 shows the socio-demographic characteristics and travel habits information, collected with the questionnaire.

|     | **Variable**     | **Answers**                                                                                                    |
| --- | ---------------- | -------------------------------------------------------------------------------------------------------------- |
| 1.  | Gender           | A. Male; B. Female                                                                                             |
| 2.  | H_income         | A. <15; B. 15-25; C. 25-35; D. 35-50; E. >50                                                                   |
| 3.  | Mother_edu       | A. Middle school diploma; B. High school diploma; C. Bachelor's degree; D. Master's degree; E. PhD or second level master |
| 4.  | Father_edu       | A. Middle school diploma; B. High school diploma; C. Bachelor's degree; D. Master's degree; E. PhD or second level master |
| 5.  | Poli_or          | A. Extreme left; B. Center-left; C. Center; D. Center-right; E. Extreme right                                  |
| 6.  | N_trav           | A. At most one; B. Two/three times; C. More than three                                                         |
| 7.  | Parents_trav     | A. At most one; B. Two/three times; C. More than three                                                         |
| 8.  | Book.sus         | A. No; B. Yes                                                                                                  |
| 9.  | Sus.accomodation | A. No; B. Yes; C. Don't know                                                                                   |
| 10. | Premium.to.sus   | A. No; B. Yes; C. Depends on amount                                                                            |
| 11. | Sus.Impo_general | A. Not at all; B. Little; C. Fairly; D. Very much                                                              |
| 12. | Sus.Impo_tourism | A. Not at all; B. Little; C. Fairly; D. Very much                                                              |
| 13. | Sus.Impo_future  | A. Not at all; B. Little; C. Fairly; D. Very much                                                              |
| 14. | Pay.will         | A. Yes, sure; B. Yes, but not happily; C. Not at all                                                           |

**Table 1:** Variables collected with the survey. Variable "H_income" is in €/thousands. Variables "N_trav" and "Parents_trav" are in average/year.

The respondent group was almost equally distributed by sex, where males constituting 57% of the sample. Household income ("H_income") serves as a pivotal indicator of young individuals' spending capacity, particularly as many are students with limited personal income. The average household income among participants is approximately €46,000 annually, exceeding the Italian national average for 2023 (€43,800, according to ISTAT data). This discrepancy may be attributed to several factors, notably the regional distribution, with 72% of participants hailing from Northern Italy, known for its higher average income compared to other regions (ISTAT, 2023). Regarding parental education levels ("Mother_edu" and "Father_edu"), most families have at least one parent with a high school diploma (53%), while 21% have attained a university degree, with a higher percentage of maternal graduates compared to paternal. Less than 5% of parents hold advanced degrees such as Ph.D. or second level master's degree. The political leanings ("Poli_or") of participants were also examined, with approximately 30% identifying as politically centrist, while 46% lean towards right-wing preferences and 25% towards left-wing preferences. This distribution mirrors Italy's general political landscape in 2022, where the center-right faction secured a majority preference of 44% in the general election. Regarding travel habits ("N_trav"), 51% of respondents reported taking between 2 and 3 holiday trips per year, while only 12% take more than 3 trips annually. The remaining 38% take at most one holiday per year. This exploratory analysis reveals a considerable degree of variability among the characteristics of young individuals within the sample. Respondents were called also to give information about the travel habits of their parents. Specifically, 41% report at most one trip per year made by their parents, while only 18% take more than 3 trips annually.

In addition, three other collected variables were considered alongside the travel-related variables. Only 31% of respondents reveal to book sustainable travel programs ("Book.sus") and only 11% choose to stay in sustainable travel accommodations ("Sus.accomodation"). Moreover, 55% of respondents that claim to do not know if the chosen accommodation adheres to sustainable programs. It is evident that while most sampled young individuals do not prioritize sustainable travel programs when planning their holidays, respondents displayed a certain degree of uncertainty in willing to pay a premium ("Premium.to.sus") to mitigate their environmental impact. In fact, 16% of respondents support this possibility and 62% claim that it depends on the additional amount to pay. Collected data suggests a potential mismatch between the sustainable initiatives promoted by industry stakeholders and their actual effectiveness in implementation. The data suggests that young tourists may not actively think about sustainable travel programs, but such initiatives are becoming more available on platforms like Booking.com. Despite this, they are still willing to reduce their impact in other ways, such as by paying a premium price. This finding holds significant implications for policymakers and industry operators in terms of strategic planning and communication. Furthermore, over half of the respondents express uncertainty regarding whether they have ever stayed in sustainable facilities, underscoring the potential limitations of current promotional and communicational strategies in driving demand or fostering the ecological transition of the tourism sector. Other variables collected concern young tourists' perception over environmental sustainability in general ("Sus.Impo_general"), for tourism industry ("Sus.Impo_tourism"), and for future investments in tourism sustainability ("Sus.Impo_future"). Responses to these questions were built on 4-levels Likert scale, according to degree of importance arranged to the sentences. Overall, the majority of young individuals attribute significance to environmental sustainability, both in general (57.6% of "Very much" and 35.3% of "Fairly") and specifically within the tourism sector (71.2% of "Very much" and 23.5% of "Fairly"). Furthermore, there is a notable acknowledgment of the necessity for future investments to enhance the environmental performance of the tourism industry (61.8% of "Very much" and 27.6% of "Fairly").

These findings align with prior research that underscores the heightened attention and awareness among younger generations towards environmental issues (e.g., Pinho and Gomes, 2023). Furthermore, it is observed that the standard deviations for general sustainability (0.73) and future sustainability (0.79) are greater than those recorded for the importance of sustainability in tourism (0.60). This suggests a higher variability and underscores the attention young people seem to place on sustainability within the tourism sector.

The questionnaire also proposed a last question concerning whether respondent for a flight priced at €140 would be willing to spend 7% more (so €10) to offset their $CO_2$ emissions generated by the flight. The idea behind the building of this variable ("Pay.will") was dictated by to test respondents' reaction in front of a real case, in which the effort for sustainability was measurable. The percentage of respondents agreeing to pay to compensate emissions is equal to 47%, while 25% would pay the fee but not happily and 28% would not pay at all. It is clearly stated that less than half of respondents is willing to do a direct effort in reducing the impact of their trips on the environment.

## 5. Results

The aim of this study is to pinpoint any distinctive characteristics within the sampled young population that differentiate those who are most attentive to environmental sustainability issues in tourism and are inclined to take proactive measures to mitigate the impact of their holidays.

The subsequent sub-sections will detail the outcomes of the LCA and the association study conducted via $\chi^2$ test to examine the relationship between the identified clusters and the "Pay.will" variable. In Appendix 1, to check robustness of the method, the same analysis was conducted by means of $K$-means algorithms.

## 5.1 Cluster analysis

The application of LCA facilitates the identification of data-driven segments based on posterior probabilities. In this study, LCA was applied to determine the optimal number of segments. Results in terms of Bayesian information criterion (BIC), Akaike Information Criterion (AIC), Log-Likelihood test, non-parametric test (Npar) and degree of freedom (DF) are reported in Table 2.

|                | 2 Classes | 3 Classes | 4 Classes |
|----------------|-----------|-----------|-----------|
| AIC            | 1688.69   | 1657.55   | 1650.32   |
| BIC            | 1779.62   | 1795.52   | 1835.33   |
| Log-Likelihood | -815.34   | -784.77   | -766.16   |
| Npar           | 29.00     | 44.00     | 59.00     |
| DF             | 141.00    | 126.00    | 111.00    |

**Table 2:** Selection of the best number of segments, ranging from 2 to 4 clusters.

As suggested by previous studies (e.g., Vermunt and Magidson, 2002), basing on results of the BIC, the optimal fit is determined by 2 clusters. Hence, the plot for the identified two clusters is presented in Figure 1.

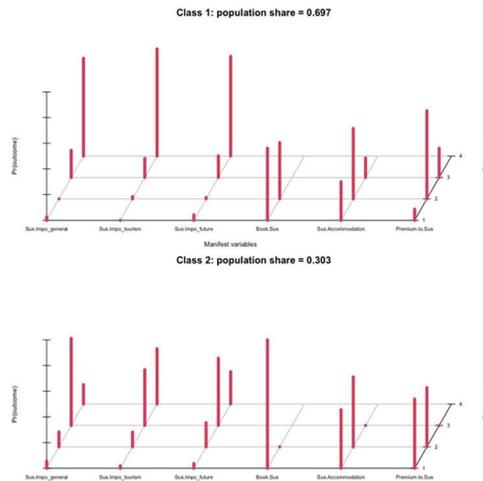

**Figure 1**: Plot of LCA variables. "Sus.Impo_general", "Sus.Impo_tourism" and "Sus.Impo_future" are on 4 levels. Variable "Book.sus" is on 2 levels. Variables "Sus.accomodation" and "Premium.to.sus" are on 3 levels.

Based on this evidence, Class 1 (comprising 69.7% of the sample) was labeled *Green Advocates* (GA), as individuals in this group actively promote sustainability, demonstrate strong awareness of environmental issues, and exhibit a pronounced willingness to undertake green actions to reduce the environmental impact of their holiday. Class 2 (comprising 30.3% of the sample) was termed *Green Learners* (GL), as individuals in this group also display awareness of the importance of environmental sustainability in tourism but are still in a learning phase regarding sustainable practices. Additionally, they show less firmness in expressing willingness to pay a premium to reduce the footprint of their tourism activities. Examining the LCA plot, it is evident that GA members tend to provide higher ratings for the variables assessing the importance of sustainability in general, in tourism and in future investments compared to GL members, who tend to respond more conservatively. Regarding sustainability variables for overnight accommodation, there is a notable difference between GA and GL. While half of the firsts consider sustainability when booking accommodation, none of the seconds do. This disparity is also reflected in their previous experiences, with none of the GL claiming to have stayed in sustainable facilities, whereas some GA do. Furthermore, the willingness to pay a premium for sustainability differs between the two clusters. In cluster 1, 25% of young people declare themselves willing to pay extra for green actions, compared to 0% in cluster 2, where individuals are divided between those not willing to mitigate their ecological footprint with a higher outlay and those who are uncertain.

Table 3 shows the sizes of the clusters and the average scores (i.e., mean values of variables for units assigned to clusters) of both descriptive and cluster variables for each segment.

|  | LCA Cluster1 $n = 119$ | LCA Cluster2 $n = 51$ |
|---|---|---|
| *Descriptive variables* | | |
| Gender | 1.44 | 1.40 |
| H_income | 3.34 | 3.25 |
| Mother_edu | 2.30 | 2.38 |
| Father_edu | 2.21 | 2.23 |
| Poli_or | 3.07 | 3.58 |
| N_trav | 1.79 | 1.62 |
| Parents_trav | 2.74 | 2.55 |
| *Cluster variables* | | |
| Book.sus | 1.44 | 1.00 |
| Sus.accomodation | 1.85 | 1.57 |
| Premium.to.sus | 2.15 | 1.47 |
| Sus.Impo_general | 3.74 | 2.89 |
| Sus.Impo_turism | 3.85 | 3.23 |
| Sus.Impo_future | 3.70 | 2.96 |

**Table 3:** Average scores of variables with LCA. Descriptive variables and clustering variables are shown.

Additionally, Table 4 displays percentage frequencies for descriptive variables in the two defined clusters.

| | GA | GL |
|---|---|---|
| *Gender* | | |
| Males | 56 | 60 |
| Females | 44 | 40 |
| *H_income* | | |
| <15 | 11 | 15 |
| 15-25 | 19 | 17 |
| 25-35 | 22 | 25 |
| 35-50 | 21 | 15 |
| >50 | 27 | 28 |
| *Mother_edu* | | |
| Middle school diploma | 20.5 | 11.3 |
| High school diploma | 52.1 | 62.3 |
| Bachelor's degree | 8.5 | 7.5 |
| Master's degree | 14.5 | 15.5 |
| PhD or second level master | 4.3 | 3.8 |
| *Father_edu* | | |
| Middle school diploma | 27.4 | 22.6 |
| High school diploma | 47.9 | 56.6 |
| Bachelor's degree | 6.8 | 1.9 |
| Master's degree | 12.8 | 13.2 |
| PhD or second level master | 5.1 | 5.7 |
| *Poli_or* | | |
| Extreme left | 3.42 | 1.89 |
| Center-left | 29.06 | 7.55 |
| Center | 27.35 | 32.08 |
| Center-right | 37.61 | 47.17 |
| Extreme right | 2.56 | 11.32 |
| *N_trav* | | |
| At most one | 32 | 49 |
| Two/three times | 56 | 40 |
| More than three | 12 | 11 |
| *Parents_trav* | | |
| No travels | 6 | 13 |
| At most one | 34 | 38 |
| Two/three times | 41 | 34 |
| More than three | 19 | 15 |

**Table 4:** Percentage distribution of descriptive variables for the 2 defined clusters.

GA members have a slightly higher average household income. Indeed, while the average household income appears similar between the two clusters, there are notable differences when examining the distribution of observations across income categories. In the GL cluster, extreme values are more frequent compared to the GA cluster. Specifically, the lowest income bracket (<€15,000/year) is reported in 15% of cases for GL, compared to 9% for GA. Regarding the level of parental education, there are no significant differences between the two segments. Travel habits also exhibit slight discrepancies between GA and GL. GA tend to travel more frequently, and according to their responses, their families also undertake more trips per year on average than GL. This may indicate greater exposure to travel experiences, potentially leading to increased awareness of tourism-related pollution issues. Lastly, another noteworthy variable showing different average scores between the segments is political orientation, which consists of five categories ("Extreme left", "Center-left", "Center", "Center-right", "Extreme right"). While the average score for GL is 3.6 out of 5, it is 3.1 for GA. Further insight into the differences between the clusters can be gleaned by breaking down the frequencies by level and computing the percentages per column, as shown in Table 4.

Values highlight notable differences in the political orientation of the subjects between the two segments. GA predominantly expresses a central political orientation, with fewer individuals at the extremes (i.e., extreme-right or extreme-left). Conversely, the GL exhibits greater polarization, with over 59% of individuals defining themselves as right-wing and less than 10% expressing left-wing political ideals. This suggests that as political polarization among young people increases, there is a corresponding decrease in attention to environmental issues, both in general and within the realm of tourism, leading to a reduced likelihood of engaging in green actions.

The final aspect of this study is to assess whether the two identified segments exhibit different propensities to adopt pro-environmental behavior.

## 5.2 $\chi^2$ test

To ascertain whether there were statistically significant differences in the profile of responses to the research question between GA and GL, a $\chi^2$ test on the variable "Pay.will" was conducted. The null hypothesis formulated considered that the profile of responses to question 15 were identical for GA and GL. The results are presented in Table 5.

|  | LCA_GA | LCA_GL |
|---|---|---|
| Very much agree | 67 | 13 |
| Little agree | 32 | 11 |
| Not at all agree | 18 | 29 |
| $n$ | 117 | 53 |
| $\chi^2$ |  | 29 |
| $df$ |  | 2 |
| $p-value$ |  | <0.001 |

**Table 5:** $\chi^2$ test for the profile of "Pay.will" variable in the two segments identified by LCA.

The $\chi^2$ test, with a statistic exceeding 29 for the segments identified by LCA, indicates that the null hypothesis of no difference between the profiles of the two groups has to be rejected. In simpler terms, the data provide high empirical evidence to suggest that GA and GL exhibit different green behavior (i.e., real willingness to pay a surcharge to offset flight $CO_2$ emissions). This conclusion is supported by the highly significant $p-values$ associated with the test (<0.001). Upon examining the relative values in Table 5, it becomes clearer that most young people in the GA segment (57%) are inclined to take green actions. Conversely, in the GL segment, most of the respondents express clear unwillingness to pay more to mitigate environmental impact (55%).

## 6.  Discussion

These findings are noteworthy for several reasons. Primarily, this research offers insight into segmenting young tourists based on their likelihood of taking tangible steps to reduce the environmental impact of tourism, independently of socio-demographic characteristics or motivations. The latent classes of GA and GL were distinguished by the subjects' sensitivity to the perceived importance of environmental sustainability issues and travel-related characteristics, such as the inclination to book more sustainable accommodations. Among all the socio-demographic characteristics examined, political orientation emerged as the most explanatory factor for differences between subjects in the clusters. The data revealed that GL exhibit a much more polarized political orientation than GA. While political orientation may not serve as a comprehensive predictor for inducing more sustainable tourism consumption, it does offer a succinct measure with strong informative power. Another intriguing aspect pertains to the discrepancy between young people uncertainty about whether they have previously stayed at sustainable facilities and their willingness to book at such facilities in the future. While neither GA nor GL can definitively recall if they have utilized green accommodation services in the past, a notable proportion of GA express interest in making reservations at sustainable facilities. In contrast, no GL report having done so. This incongruity emphasizes the importance for policymakers and practitioners to bridge the gap between willingness and experience (or recollection of experience). In other words, there seems to be a gap between what young people would be willing to do (i.e., book at eco-friendly accommodation) and what they remember doing in past holidays (i.e., knowing they ever stayed at sustainable accommodation). This knowledge gap may stem from two primary factors. Firstly, it may be an identification issue, wherein young people find it challenging to discern the most sustainable housing when booking. In this scenario, policymakers could intervene by implementing appropriate and standardized communication measures for hosting facilities, allocating funding to incentivize more establishments to adopt eco-friendly practices, or encouraging participation in independent sustainability programs such as the Green Key and BioHotels platforms.

On the other hand, it may be the hosts who need to take action to improve the perception of sustainability of their facility. An accommodation, even if this is not explicitly part of green programs, but the property has invested in sustainable strategies, should make this information explicit to the tourist. In this way, the visitor would have a more vivid reminder that he/she "stayed in an environmentally conscious facility," and this could trigger subsequent re-adoptions or approaching the environmental issue. Indeed, this analysis underscores the necessity for stakeholders in the tourism sector to delve deeper into these issues and implement concrete efforts to enhance the environmental sustainability of tourism. This entails addressing both communication and experiential aspects, whether in the booking process or on-site, to foster greater awareness and action regarding environmental impacts. Another crucial finding from this analysis is the statistically significant difference between GA and GL regarding their propensity to mitigate their environmental footprint through increased expenditure. This divergence remains significant even when considering income conditions, indicating that the willingness to pay more for sustainability is driven by factors beyond financial means. Moreover, the distinction between perceiving sustainability as "fairly important" versus "very much important" emerges as a significant discriminating factor. This differentiation delineates between those who are merely aware of the environmental impact of tourism (the GL) and those who, in addition to awareness, demonstrate a proactive willingness to address such pollution (the GA).

## 7. Conclusion

This study delves into the crucial intersection of tourism and environmental sustainability, exploring the pro-environmental behaviors of young tourists. Against the backdrop of an expanding tourism sector, concerns about its environmental impact have gained prominence. While tourism contributes significantly to global GDP and job creation, it also poses challenges to environmental preservation. By leveraging latent class cluster analysis this research identifies two distinct segments among young tourists: Green Advocates (GA) and Green Learners (GL). These segments exhibit divergent attitudes toward environmental sustainability, with GA demonstrating a stronger commitment to green actions compared to GL. Key insights emerge from the analysis of variables associated with sustainability, accommodation choices, and willingness to pay a premium for eco-friendly practices. Notably, political orientation emerges as a significant explanatory factor for differences between the segments, with GL exhibiting greater political polarization compared to GA. Moreover, a noteworthy finding is a discrepancy between young people's willingness to book sustainable accommodations and their recollections of past experiences. This highlights a knowledge gap that policymakers and practitioners must address through improved communication and experiential initiatives. The study's findings have practical implications for DMOs and policymakers, offering insights to inform decision-making processes and communication strategies aimed at promoting environmental sustainability in the tourism sector. Furthermore, the research contributes to the literature by providing empirical evidence and identifying distinct segments within the young tourist demographic. Moving forward, stakeholders in the tourism sector must deepen their understanding of these issues and implement concrete efforts to enhance environmental sustainability. Addressing the knowledge gap and fostering greater awareness and action among young tourists are critical steps toward achieving sustainable tourism practices. Additionally, recognizing the significance of factors beyond financial means in driving pro-environmental behaviors underscores the need for multifaceted strategies to promote sustainability in the tourism sector.

## Declaration of interest

The authors declare that they have no known competing financial interests or personal relationships that could have appeared to influence the work reported in this article.

## Author contributions

The authors equally contributed to the work.

## Data availability

Data used in the paper were collected by means of a survey on Italian students. Participants give their information anonymously, in the respect of privacy rules.

# References

Ashraf, M. S., F. Hou, W. G. Kim, W. Ahmad, and R. U. Ashraf (2020). Modeling tourists' visiting intentions toward ecofriendly destinations: Implications for sustainable tourism operators. *Business Strategy and the Environment,* 29(1), 54– 71.

Barr, S., G. Shaw, and T. Coles (2011). Times for (Un) sustainability? Challenges and opportunities for developing behaviour change policy. A case-study of consumers at home and away. *Global Environmental Change,* 21(4), 1234–1244.

Beall, J. and B. B. Boley (2022). An ecotourist by whose standards? Developing and testing the Ecotourist Identification Scale (EIS). *Journal of Ecotourism,* 21(2), 99– 120.

Blamey, R. K. (2001). *Principles of ecotourism*, The encyclopedia of ecotourism. New York: CAB International.

Bridges, C. M. and W. B. Wilhelm (2008). Going beyond green: The "why and how" of integrating sustainability into the marketing curriculum. *Journal of marketing education,* 30(1), 33–46.

Buffa, F. (2015). Young tourists and sustainability. Profiles, attitudes, and implications for destination strategies. *Sustainability,* 7(10), 14042–14062.

Butler, R. W. (1999). Sustainable tourism: A state-of-the-art review. *Tourism geographies,* 1(1), 7–25.

DeSarbo, W. S. and C. F. DeSarbo (2001). A generalized normative segmentation methodology employing conjoint analysis. In *Conjoint measurement: methods and applications* (pp. 473-504). Berlin, Heidelberg: Springer Berlin Heidelberg.

Dolnicar, S. (2004). Beyond "commonsense segmentation": A systematics of segmentation approaches in tourism. *Journal of Travel Research,* 42(3), 244–250.

Dolnicar, S. and K. Matus (2008). Are green tourists a managerially useful target segment?. *Journal of Hospitality & Leisure Marketing,* 17(3-4), 314–334.

ETIS (2017). *The European Tourism Indicator System: ETIS toolkit for sustainable destination management*.

Everitt, B. S. (1988). A finite mixture model for the clustering of mixed-mode data. *Statistics & probability letters,* 6(5), 305–309.

Everitt, B. S., S. Landau, and M. Leese (1992). *Cluster Analysis*. Edward Arnold.

Fayos-Solà, E., M. D. Alvarez, and C. Cooper (2014). *Tourism as an instrument for development: A theoretical and practical study*. Emerald Group Publishing.

Federation, E. (2010). *European Charter for Sustainable Tourism in Protected Areas*.

Fraley, C. and A. E. Raftery (1998). How many clusters? Which clustering method? Answers via model-based cluster analysis. *The computer journal,* 41(8), 578–588.

Fredman, P., S. Wall-Reinius, and A. Grundén (2012). The nature of nature in naturebased tourism. *Scandinavian Journal of Hospitality and Tourism,* 12(4), 289–309.

Gardyn, R. (2002). Educated consumers. *American Demographics,* 24(10), 18–19.

Garvey, A. M. and L. E. Bolton (2017). Eco-product choice cuts both ways: How proenvironmental licensing versus reinforcement is contingent on environmental consciousness. *Journal of Public Policy & Marketing,* 36(2), 284–298.

Gillison, F. B., P. Rouse, M. Standage, S. J. Sebire, and R. M. Ryan (2019). A metaanalysis of techniques to promote motivation for health behaviour change from a self-determination theory perspective. *Health psychology review,* 13(1), 110–130.

Goodwin, H. (2016). *Responsible tourism: Using tourism for sustainable development*. Goodfellow Publishers Ltd.

Han, H. (2021). Consumer behavior and environmental sustainability in tourism and hospitality: A review of theories, concepts, and latest research. *Sustainable Consumer Behaviour and the Environment*, 1–22.

Jennings, G., C. Cater, L. Y. Lee YoungSook, C. Ollenburg, A. Ayling, and B. Lunny (2009). Generation Y: perspectives of quality in youth adventure travel experiences in an Australian backpacker context. In *Tourism and generation Y* (pp. 58-72). Wallingford UK: Cabi.

Kamakura, W. A. and M. Wedel (1995). Life-style segmentation with tailored interviewing. *Journal of Marketing Research*, 32(3), 308–317.

Khanra, S., A. Dhir, P. Kaur, and M. Mäntymäki (2021). Bibliometric analysis and literature review of ecotourism: Toward sustainable development. *Tourism Management Perspectives ,*37, 100777.

Kim, J., C. R. Taylor, K. H. Kim, and K. H. Lee (2015). Measures of perceived sustainability. *Journal of Global Scholars of Marketing Science,* 25(2), 182–193.

Leask, A., A. Fyall, and P. Barron (2015). *Generation Y: Market Opportunity or Marketing Challenge–Strategies to Engage Generation Y in the UK Attractions' Sector'*.

Lu, J. and S. K. Nepal (2009). Sustainable tourism research: An analysis of papers published in the Journal of Sustainable Tourism. *Journal of sustainable Tourism,* 17(1), 5–16.

Madhavan, H. and R. Rastogi (2013). Social and psychological factors influencing destination preferences of domestic tourists in India. *Leisure Studies,* 32(2), 207– 217.

Massidda, C. and I. Etzo (2012). The determinants of Italian domestic tourism: A panel data analysis. *Tourism Management,* 33(3), 603–610.

Moscardo, G., L. Murphy, and P. Beckendorff (2011). *Generation Y and travel futures*. Goodfellow Publishers.

Pendergast, D. (2009). Getting to know the Y generation. In *Tourism and generation Y* (pp. 1-15). Wallingford UK: Cabi.

Pinho, M. and S. Gomes (2023). Generation Z as a critical question mark for sustainable tourism–An exploratory study in Portugal. *Journal of Tourism Futures,* ahead-of-print.

Reinikainen, H., J. T. Kari, and V. Luoma-Aho (2020). Generation Z and organizational listening on social media. *Media and Communication,* 8(2), 185–196.

Rigdon, E. E., C. M. Ringle, and M. Sarstedt (2010). Structural modeling of heterogeneous data with partial least squares. *Review of marketing research*, 255–296.

Rondan-Cataluna, F. J., M. J. Sanchez-Franco, and A. F. Villarejo-Ramos (2010). Searching for latent class segments in technological services. *The Service Industries Journal,* 30(6), 831–849.

Sánchez-Fernández, R., M. Á. Iniesta-Bonillo, and A. Cervera-Taulet (2019). Exploring the concept of perceived sustainability at tourist destinations: A market segmentation approach. *Journal of Travel & Tourism Marketing,* 36(2), 176–190.

Schwartz, S. H. (1994). Are there universal aspects in the structure and contents of human values?. *Journal of social issues,* 50(4), 19–45.

Streimikiene, D., B. Svagzdiene, E. Jasinskas, and A. Simanavicius (2021). Sustainable tourism development and competitiveness: The systematic literature review. *Sustainable development,* 29(1), 259–271.

Vermunt, J. K. and J. Magidson (2002). Latent class cluster analysis. *Applied latent class analysis,* 11(89-106), 60.

Vukic, M., M. Kuzmanovic, and M. Kostic Stankovic (2015). Understanding the heterogeneity of Generation Y's preferences for travelling: A conjoint analysis approach. *International Journal of Tourism Research,* 17(5), 482–491.

Wang, J., S. Wang, H. Xue, Y. Wang, and J. Li (2018). Green image and consumers' wordof-mouth intention in the green hotel industry: The moderating effect of Millennials. *Journal of cleaner production,* 181, 426–436.

Wang, S., J. Wang, J. Li, and F. Yang (2020). Do motivations contribute to local residents' engagement in pro-environmental behaviors? Resident-destination relationship and pro-environmental climate perspective. *Journal of Sustainable Tourism,* 28(6), 834– 852.

Weaver, D. B. (2014). Asymmetrical dialectics of sustainable tourism: Toward enlightened mass tourism. *Journal of Travel Research,* 53(2), 131–140.

Weaver, D. B. and L. J. Lawton (2007). Twenty years on: The state of contemporary ecotourism research. *Tourism management,* 28(5), 1168–1179.

Wedel, M. and W. S. DeSarbo (1994). A review of recent developments in latent class regression models. *Advanced methods of marketing research*, 352–388.

Wedel, M. and W. A. Kamakura (2000). *Market segmentation: Conceptual and methodological foundations*. Springer Science & Business Media.

Williams, K. C., R. A. Page, A. R. Petrosky, and E. H. Hernandez (2010). Multi-generational marketing: Descriptions, characteristics, lifestyles, and attitudes. *The journal of applied business and economics,* 11(2), 21.

WTTC (2022). *Travel tourism: Economic impact 2021*. World Travel and Tourism Council.

Wu, J., X. Font, and J. Liu (2021). Tourists' pro-environmental behaviors: Moral obligation or disengagement?. *Journal of Travel Research,* 60(4), 735–748.

Xu, B. and D. Dong (2020). Evaluating the impact of air pollution on China's inbound tourism: A gravity model approach. *Sustainability,* 12(4), 1456.

Yeoman, I., C. Hsu, K. Smith, and S. Watson (2010). *Tourism and demography*. Goodfellow Publishers Ltd.

# Appendix 1

In the $K$-means algorithm, centroids are initially randomly initialized using observations in the dataset. Observations are then assigned to the cluster with the closest centroid, typically determined by a chosen distance measure such as Euclidean distance. Centroids are updated iteratively based on the mean of observations assigned to each cluster. These operations are repeated iteratively until the centroids converge or until a maximum number of iterations is reached. $K$-means necessitates data normalization before implementation and often relies on principal components for graphical representation, which may compromise interpretative simplicity. Nevertheless, analyzing the size of the first two components in the $X$ and $Y$ axes can provide insights into the primary directions of data variation.

In the present case, the $K$-means algorithm has been implemented to model observations into 2 segments. The inclusion of $K$-means serves the purpose of comparing results between two different algorithms, thereby enhancing the robustness of the estimates.

In the following, results in terms of plot (Figure A1), average scores of variables (Table A1) and $\chi^2$ test (Table A2) are reported.

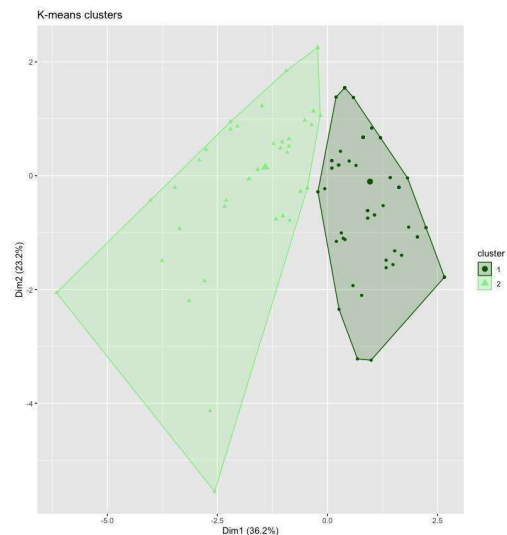

**Figure A1:** Plot of $K$-means clustering algorithm. Centroids are reported (a principal component analysis was computed on the normalized data to get the $Y$ and $X$ dimensions). Variables "Sus.Impo_general", "Sus.Impo_tourism" and "Sus.Impo_future" are on 4 levels. Variable "Book.sus" is on 2 levels. Variables "Sus.accomodation" and "Premium.to.sus" are on 3 levels.

|  | K-Means Cluster1 n = 101 | K-Means Cluster2 n = 69 |
|---|---|---|
| *Descriptive variables* | | |
| Gender | 1.46 | 1.39 |
| H_income | 3.33 | 3.29 |
| Mother_edu | 2.28 | 2.39 |
| Father_edu | 2.21 | 2.22 |
| Poli_or | 3.05 | 3.49 |
| N_trav | 1.80 | 1.65 |
| Parents_trav | 2.76 | 2.55 |
| *Cluster variables* | | |
| Book.sus | 1.50 | 1.01 |
| Sus.accomodation | 1.93 | 1.51 |
| Premium.to.sus | 2.20 | 1.55 |
| Sus.Impo_general | 3.82 | 2.96 |
| Sus.Impo_turism | 3.85 | 3.36 |
| Sus.Impo_future | 3.79 | 3.00 |

**Table A1:** Average scores of variables with *K*-means. Descriptive variables and clustering variables are shown.

|  | K-means_GA | K-means_GL |
|---|---|---|
| Very much agree | 62 | 18 |
| Little agree | 27 | 16 |
| Not at all agree | 12 | 35 |
| $n$ | 101 | 69 |
| $\chi^2$ | | 33 |
| $df$ | | 2 |
| $p-value$ | | <0.001 |

**Table A2**: $\chi^2$ test for the profile of "Pay.will" variable in the two segments identified by *K*-means.